\documentclass [aps,prd,eqsecnum,showpacs,superscriptaddress,twocolumn,floatfix,preprintnumbers,altaffilletter]{revtex4}

\usepackage{graphicx}
\usepackage{amssymb}

\begin{document}

\title{Opportunity to Test non-Newtonian Gravity Using Interferometric Sensors with Dynamic Gravity Field Generators}

\author{P\'{e}ter~Raffai}
\email{praffai@bolyai.elte.hu}
\affiliation{E\"otv\"os University, Institute of Physics, 1117 Budapest, Hungary}

\author{G\'{a}bor~Szeifert}
\affiliation{E\"otv\"os University, Institute of Physics, 1117 Budapest, Hungary}

\author{Luca~Matone}
\affiliation{Columbia University, Department of Physics, New York, NY 10027, USA}

\author{Yoichi Aso}
\affiliation{The University of Tokyo, Department of Physics, Tokyo, 113-0033, Japan}

\author{Imre~Bartos}
\affiliation{Columbia University, Department of Physics, New York, NY 10027, USA}

\author{Zsuzsa~M\'{a}rka}
\affiliation{Columbia University, Department of Physics, New York, NY 10027, USA}

\author{Fulvio~Ricci}
\affiliation{Dipartimento di Fisica, Universit\`{a} La Sapienza, Rome, I-00185, Italy,\\ INFN Sezione di Roma, Rome, I-00185, Italy}

\author{Szabolcs~M\'{a}rka}
\affiliation{Columbia University, Department of Physics, New York, NY 10027, USA}

\begin{abstract}
We present an experimental opportunity for the future to measure possible violations to Newton's $1/r^2$ law in the $0.1-10\ \mathrm{meter}$ range using Dynamic gravity Field Generators (DFG) and taking advantage of the exceptional sensitivity of modern interferometric techniques. The placement of a DFG in proximity to one of the interferometer's suspended test masses generates a change in the local gravitational field that can be measured at a high signal to noise ratio. The use of multiple DFGs in a null experiment configuration allows to test composition independent non-Newtonian gravity significantly beyond the present limits. Advanced and third-generation gravitational-wave detectors are representing the state-of-the-art in interferometric distance measurement today, therefore we illustrate the method through their sensitivity to emphasize the possible scientific reach. Nevertheless, it is expected that due to the technical details of gravitational-wave detectors, DFGs shall likely require dedicated custom configured interferometry. However, the sensitivity measure we derive is a solid baseline indicating that it is feasible to consider probing orders of magnitude into the pristine parameter well beyond the present experimental limits significantly cutting into the theoretical parameter space.
\end{abstract}

\maketitle

\section{Introduction}
\label{introduction}

Beyond the original goal of detecting gravitational waves, modern interferometric sensors (IFO) (such as the future Advanced LIGO (aLIGO)~\cite{adligohomepage}, the Advanced VIRGO (aVIRGO)~\cite{advirgohomepage}, the LCGT~\cite{LCGTOverview}, the AEI 10m~\cite{AEI10m} interferometer; and the proposed third generation detectors such as the Einstein Telescope (ET)~\cite{ChassandeMottinetal}) can be viewed as unique pioneering instruments capable of measuring induced displacements below the $10^{-20}~-~10^{-19}~\mathrm{m}/\sqrt{\mathrm{Hz}}$ scale~\cite{snomass2001}. This feature opens up myriads of new possibilities for fundamental science. In searches for gravitational waves (GW) the signature of the induced displacement of the IFO's test mass (TM) ranges from the basically unknown to the well-predicted but yet undetected. In contrast to this, one may design a precision experiment in which the local gravitational field around the TM of the interferometric sensor is modulated at a given frequency on a well-controlled manner by a dynamic gravity field generator (DFG). A DFG consists essentially of a rotating mass of null odd and non-null even moments, from which the quadrupole term dominates. In the Newtonian limit, its effect on the detector is to add a signal at the even-harmonics of the rotation frequency. A well-characterized DFG has the potential to provide sub-percent calibration for the GW detectors in phase and amplitude as well as to evaluate current theoretical Newtonian noise estimates. This is the subject of a separate publication~\cite{rotorCQG}.

It has been shown in the past that devices capable of producing gravity field gradients can be employed, in conjunction with a suitable displacement sensor, for testing Newton's $1/r^2$ law in the laboratory scale. In the first experiment in 1967 Forward and Miller \cite{ForwardMiller1967} used an orbiter sensor, originally developed for measuring the lunar mass distribution, to test Newton's $1/r^2$ law in the $10\ \mathrm{cm}$ scale. Weber and Sinsky \cite{SinskyWeber1967,Sinsky1968} used a GW bar detector as a sensor, acoustically stressing a volume of matter at $1660\ \mathrm{Hz}$, measuring an excess in noise in the detector consistent to theory.

In the 1980s at the University of Tokyo, a series of experiments were carried through to test violations to the inverse square law (ISL) up to a distance of $10\ \mathrm{m}$~\cite{Hirakawa1980,oide,suzuki,Ogawa1982,Kuroda1985}. In these studies, the coupling between the dynamic gravity field generated by a rotating mass, and the quadrupole moment of a mechanical oscillator antenna was measured as a function of the rotor-antenna separation. With this method limits on non-Newtonian gravity were provided and a measurement of the Newtonian constant, G, was found in agreement with previous experiments ~\cite{Ogawa1982,Kuroda1985}.

In the 1990s, the gravitational-wave group at the University of Rome developed and carried out experiments \cite{Astone1991,Astone1998} on the cryogenic GW bar detector, EXPLORER, at CERN. A device, rotating at a frequency close to half of the antenna's resonant frequency was developed, and the resulting dynamic field was measured as a function of the source-sensor separation. The results were then used to derive upper limits to Yukawa-like gravitational potential violations at laboratory scale.

The experimental designs cited above consisted in the use of a single DFG in conjunction with bar type GW detectors. In this article we propose a significantly different design, exploiting the exceptional bandwidth and sensitivity of state-of-the-art interferometric sensors. We illustrate the method through the sensitivity of advanced and third-generation gravitational wave detectors to emphasize the feasible scientific reach. However, it is expected that due to the technical details of gravitational wave detectors, the DFGs based ISL measurements shall require dedicated custom configured interferometry. Nevertheless, the sensitivity measure we derive is a solid baseline. The concept is based on a null-experiment configuration, where a \emph{pair} of well matched and symmetrical DFGs, rotating at the same frequency but $90^{\circ}$ out of phase, generates a null-signal in the absence of violations to Newton's $1/r^2$ law. In the presence of violations, and within experimental uncertainties, the effect of the two DFGs would not cancel, and a measure of such deviations would be achievable.

\begin{figure}
\begin{center}
  \includegraphics[height=3.4in,angle=-90]{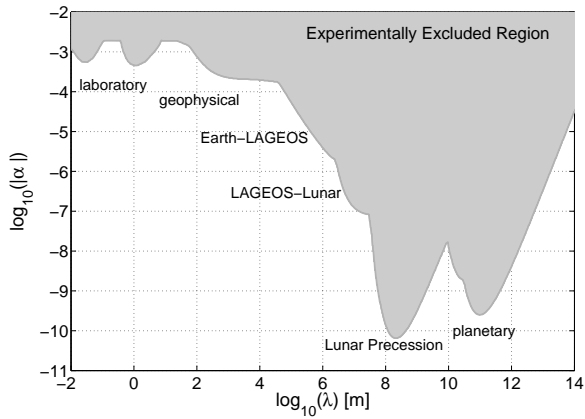}\\
  \caption{Current $95\%$-confidence-level constraints on the
  inverse-square-law-violating Yukawa interactions with $\lambda > 1\ \mathrm{cm}$.
  (Courtesy of Ref.\cite{AdelbergerReview2003}, Figure(4), with the addition
  of \cite{Hoskins1985} and \cite{MoodyPaik} data)}
  \label{summary}
\end{center}
\end{figure}

The most common way of interpreting composition independent tests of non-Newtonian gravity (for review see e.g.~\cite{AdelbergerReview2003} and ~\cite{NewmanBergBoynton2009}) is through the Yukawa formalism, where an additional so-called Yukawa term ($V^\mathrm{Y}(r)$) is added to the classical Newtonian gravitational potential ($V^\mathrm{N}(r)$):

\begin{equation}
\label{pot} V(r) = V^\mathrm{N}(r) + V^\mathrm{Y}(r) = -G ~ \frac{mM}{r} ~ [ 1 + \alpha e^{-r/\lambda} ]
\end{equation}

In this expression $G$ is the gravitational constant, $m$ and $M$ are the interacting masses, and $r$ denotes the distance between the two point masses. The Yukawa parameters of interest are $\alpha$, denoting the Yukawa interaction coupling strength, and $\lambda$, giving the length scale of the coupling. For a Yukawa range of $\lambda=0.1-10\ \mathrm{m}$, the current limit on the coupling strength, $\alpha$, is in the order of $10^{-4}-10^{-3}$ for both negative, positive, and absolute value of $\alpha$. (see Fig. \ref{summary}.). The concept designed here shows promise to explore deviations from ISL significantly below present bounds, maybe even down to the order of $10^{-6}$ if enabled by technology and cost.

In this work we study the application of future interferometric sensors for artificially generated gravity fields by a pair of hypothetical DFGs in a null experiment configuration. We show that a pair of symmetrical and matched DFGs cancel each other's effect on the TM at twice the rotation frequency in the Newtonian limit. Thus we relate the estimated test mass displacement to non-Newtonian parametrization of the law of gravitation. We also consider the contribution from some of the undesired system asymmetries due to production/measurement uncertainties. We numerically compute the tolerances of a hypothetical experimental setup to measure Yukawa-like deviations from the $1/r^2$ law and we estimate a limit on the coupling strengths, $\alpha$-s, that might be achieved with future measurements.

\section{Gravity field dynamics from a two-mass DFG}
\label{NFD}

Consider a hypothetical DFG, consisting of two point masses, $m_\mathrm{1}$ and $m_\mathrm{2}$, separated by a distance of $r_\mathrm{1}$ and $r_\mathrm{2}$ respectively from the center of rotation, rotating at a frequency of $f_\mathrm{0} = \omega_\mathrm{0}/(2 \pi)$. A point mass, $M$, representing the TM, is then placed at distance $d\ (>r_\mathrm{1,2})$ from the DFG's center of rotation.

\begin{figure}
\begin{center}
  \includegraphics[width=3.4in]{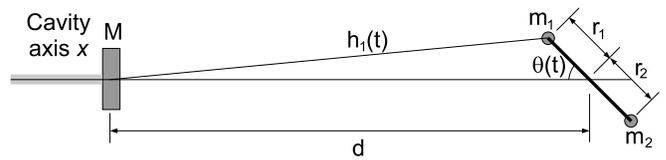}\\
  \caption{Schematic of a hypothetical asymmetric two-mass DFG configuration. Point masses $m_\mathrm{1}$ and $m_\mathrm{2}$ are rotating with a frequency of $f_\mathrm{0}$ and radii of $r_\mathrm{1}$ and $r_\mathrm{2}$ from the center of rotation, respectively. The DFG center of rotation is placed in a distance of $d$ from the mass $M$, representing the TM center of mass. The distance between the DFG masses and the TM are $h_\mathrm{1}$ and $h_\mathrm{2}$, respectively, $\theta(t)=\omega_\mathrm{0} t$, and only accelerations of the TM along the IFO cavity axis are considered.}
  \label{DFGsketch}
\end{center}
\end{figure}

We first calculate the acceleration of $M$ induced by the DFG along the axis connecting $M$ and the center of rotation of the DFG. This axis corresponds to the optical axis of the IFO. Introducing the dimensionless parameters $R_\mathrm{1} = r_\mathrm{1}/d$, and $R_\mathrm{2} = -r_\mathrm{2}/d$, the distance between the DFG's $i$-th mass and the point mass $M$, $h_\mathrm{i}$, can be written as

\begin{equation}\label{hi}
h_\mathrm{i}(t)  = d ~ \sqrt {1 + R_\mathrm{i}^2  - 2R_\mathrm{i}\cos \theta(t) }
\end{equation}

where $\theta(t)=\omega_\mathrm{0} t$. The Newtonian potential at the position of point mass $M$ is then given by

\begin{equation}
V^\mathrm{N}  = \sum\limits_{i = 1}^2 {V_\mathrm{i}^\mathrm{N} }  = - GM\sum\limits_{i = 1}^2 {\frac{{m_\mathrm{i} }} {{h_\mathrm{i}}}}.
\end{equation}

The magnitude of the induced acceleration on $M$ along the optical axis can be written as

\begin{equation}\label{cla_acc}
a^\mathrm{N}  = \frac{1} {M}\left| {\frac{{\partial V^\mathrm{N} }}
{{\partial d}}} \right| = \frac{G} {{d^2 }}\sum\limits_{i = 1}^2
{m_\mathrm{i} B_\mathrm{i} (R_\mathrm{i} ,\theta )}.
\end{equation}

Here $B_\mathrm{i} (R_\mathrm{i} ,\theta )$ is a geometrical factor

\begin{equation}
B_\mathrm{i} (R_\mathrm{i} ,\theta ) = \frac{{1 - R_\mathrm{i} \cos \theta }} {{\left( {1 +
R_\mathrm{i}^2 - 2R_\mathrm{i} \cos \theta } \right)^{{3 \mathord{\left/
 {\vphantom {3 2}} \right.
 \kern-\nulldelimiterspace} 2}} }}.
\end{equation}

As we have shown elsewhere~\cite{rotorCQG}, for the case of a much smaller lever arm $r_\mathrm{i}$ than distance $d$ ($R_\mathrm{i} \ll 1$), considering only the dominant terms in the first few harmonics, the induced displacement of a free mass $M$ along the axis connecting it to the DFG's center of rotation can be approximated as

\begin{eqnarray}
\label{displ2}
x^N(t) \simeq \frac{G}{(d ~ \omega_\mathrm{0} )^2} \times \Bigg{[} 2
\cdot \frac{\mathcal{M}_\mathrm{1}}{d} \cdot \cos{\omega_\mathrm{0}
t} &+&\\ \nonumber & & \hskip -2.0in \frac{9}{16} \cdot
\frac{\mathcal{M}_\mathrm{2}}{d^2} \cdot \cos{2 \omega_\mathrm{0} t}
+  \frac{5}{18} \cdot \frac{\mathcal{M}_\mathrm{3}}{d^3} \cdot
\cos{3 \omega_\mathrm{0} t} + \cdot\cdot\cdot \Bigg{]}
\end{eqnarray}

where $\mathcal{M}_\mathrm{n}$ describes the $n$-th multipole moment of the DFG's mass distribution. We observe that using Eq.~(\ref{displ2}) and in the case of a DFG mass distribution invariant under rotation by pi, all odd moments vanish and the induced displacement is dominated by the quadrupole moment $\mathcal{M}_\mathrm{2}$ at twice the rotation frequency. We note, that the solution represented in Eq.~(\ref{displ2}) is valid for a free body, which, in case of a suspended TM is a good approximation for frequencies well above the eigenfrequencies of the suspension (typically around $0.1 - 1\ \mathrm{Hz}$ \cite{TMsuspension}).

In the same way as above, we calculate the field dynamics arising from a Yukawa-like potential perturbation to the classical Newtonian field. In view of the fact that the potentials are additive and all operations to calculate the acceleration are linear, the acceleration of point mass $M$ due to the two potential terms will also be additive. Using Eq.~(\ref{pot}), the Yukawa perturbation to the Newtonian potential at $M$ due to the DFG can be expressed as

\begin{equation}
V^\mathrm{Y}  = \sum\limits_{i = 1}^2 {V_\mathrm{i}^\mathrm{Y}  =  -
\alpha GM\sum\limits_{i = 1}^2 {\frac{{m_\mathrm{i} }}
{{h_\mathrm{i} }}e^{{{ - h_\mathrm{i} } \mathord{\left/
 {\vphantom {{ - h_\mathrm{i} } \lambda }} \right.
 \kern-\nulldelimiterspace} \lambda }} } }
\end{equation}

where $h_\mathrm{i}=h_\mathrm{i}(t)$ is given by Eq.~(\ref{hi}). In the same way, the magnitude of the TM induced acceleration along the optical axis is

\begin{equation}\label{yuk_acc}
a^\mathrm{Y}  = \frac{1} {M}\left| {\frac{{\partial V^\mathrm{Y} }}
{{\partial d}}} \right| = \alpha ~ \frac{{G}} {{d^2 }}\sum\limits_{i
= 1}^2 {m_\mathrm{i} B_\mathrm{i} (R_\mathrm{i} ,\theta )
f_\mathrm{i} (R_\mathrm{i} ,\theta ,\lambda)}.
\end{equation}

where $\theta = \theta(t) = \omega_\mathrm{0} t$, and $f_\mathrm{i} (R_\mathrm{i} ,\theta ,\lambda)$ is a function of the length scale of the Yukawa coupling, $\lambda$:

\begin{eqnarray}
f_\mathrm{i}(R_\mathrm{i} ,\theta ,\lambda) = \Bigg{[} 1 +
\frac{d}{\lambda}\sqrt{1+R_\mathrm{i}^2-2R_\mathrm{i}\cos{\theta}} \Bigg{]} &\times& \\
\nonumber && \hskip -1.5in \times \exp{\Bigg{(}-\frac{d}{\lambda }
\sqrt{1 + R_\mathrm{i}^2 - 2R_\mathrm{i} \cos \theta}\Bigg{)}}
\end{eqnarray}

Analytical integration of Eq.~(\ref{cla_acc}) and Eq.~(\ref{yuk_acc}) cannot be easily formulated without using some additional level of approximation. For this reason, we treat the problem numerically, facilitating the treatment of some of the experimental uncertainties. We have modeled both the Newtonian and the non-Newtonian dynamics on the TM while taking into account many of the DFG fabrication and measurement procedural uncertainties.

In this work the TM is approximated as a driven, damped pendulum. Let the pendulum's longitudinal eigenfrequency be $\omega_\mathrm{p}$ with quality factor $Q$. Casting the differential equations of motion (Eq.~(\ref{cla_acc}) and Eq.~(\ref{yuk_acc})) in the Laplace frequency domain, the transfer function relating the acceleration $a^\mathrm{N,Y}$ to the induced displacement $x^\mathrm{N,Y}$ is

\begin{equation}
\label{susp}
x^\mathrm{N,Y}(s) = \frac{a^\mathrm{N,Y}(s)}{s^2 +
(\omega_\mathrm{p}/Q) s + \omega_\mathrm{p}^2}
\end{equation}

where $s=i \omega$ (the upper index $\mathrm{N,Y}$ denotes that the same expression applies to both Newtonian and Yukawa case).

We used Monte Carlo simulations to model a null-experiment design using DFGs to measure hypothetical violations to the classical inverse square law. To do this, we first numerically computed the induced acceleration due to the Newtonian and non-Newtonian terms using Eq.~(\ref{cla_acc}) and Eq.~(\ref{yuk_acc}). These results, generated in the time domain, are then mapped in the Fourier-domain to generate an acceleration spectrum, which is then filtered through the transfer function presented in Eq.~(\ref{susp}). The displacement contributions at the different harmonics, for different sets of parameters and uncertainties are then computed and analyzed.

\begin{figure}
\begin{center}
  \includegraphics[width=3.4in]{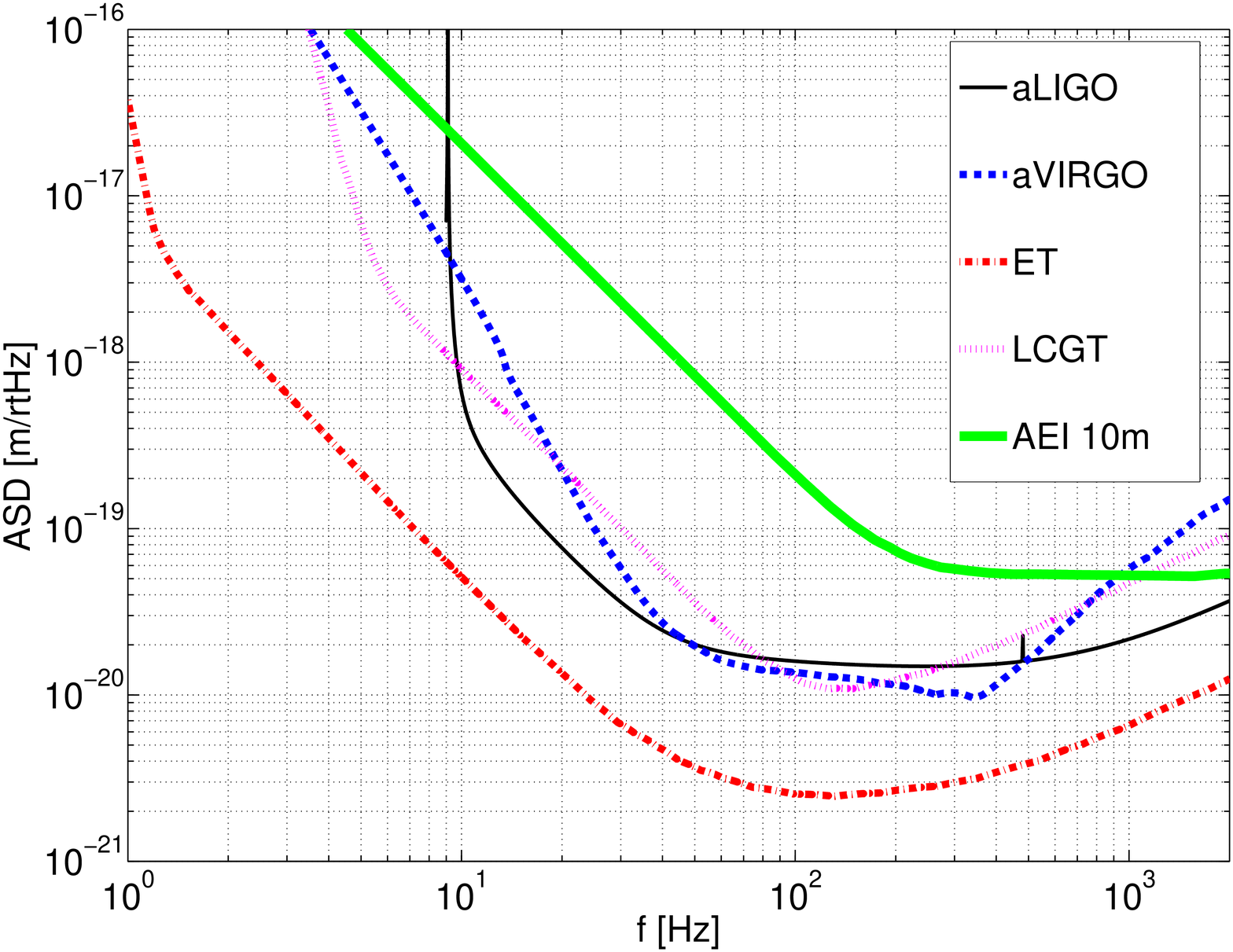}\\
  \caption{The design displacement sensitivity of aLIGO (thin black), aVIRGO (dashed), the ET (dot-dashed), LCGT (dotted), and AEI 10m (thick green).}
  \label{SensCurves}
\end{center}
\end{figure}

To assess the short and long term feasibility of the proposed method, the simulated results are compared to the displacement sensitivity of future most sensitive interferometric GW detectors. Fig.~\ref{SensCurves}. shows the design sensitivities for aLIGO, aVIRGO, the ET (for a reference to these, see~\cite{ChassandeMottinetal}), LCGT~\cite{LCGTOverview}, and the AEI 10m detector~\cite{AEI10m}.

\section{Hypothetical Inverse Square law violation measurements with DFGs}
\label{yuk}

\begin{figure}
\begin{center}
  \includegraphics[width=3.4in]{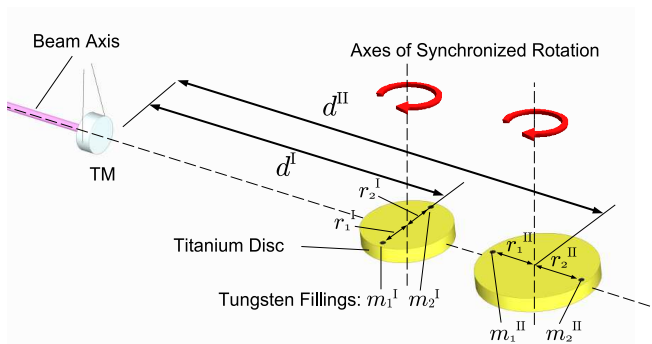}\\
  \caption{Null experiment geometry for the measurement of a possible  deviation from Newton's $1/r^{2}$ law. Two conceptualized DFGs are placed in-line with one of the IFO's arms at a distance of $d^\mathrm{I}$ and $d^\mathrm{II}$ from the TM. The DFGs positions and quadrupole moments are chosen such that when operated at a relative phase of $\beta=\pi/2$ the net displacement at $2 f_\mathrm{0}$ will largely depend on the non-Newtonian coupling strength $\alpha$.}
  \label{tworotorD}
\end{center}
\end{figure}

In order to mitigate the effect of the IFO's calibration uncertainty, two DFGs rotating at the same frequency ($f_\mathrm{0}$), but out of phase ($\beta = \pi/2$), are set up to generate a null effect on the TM in the Newtonian limit at twice the rotation frequency. For this reason, the uncertainties of the proposed measurement, whose setup is shown in Fig.~\ref{tworotorD}., will solely depend on the production and procedural uncertainties of DFGs.

For example, similarly to the one proposed in~\cite{rotorCQG} as a detector calibration tool, a single hypothetical DFG may consist of an Aircraft Grade (6Al/6V/2Sn) Titanium disc $60\ \mathrm{cm}$ in diameter and $10\ \mathrm{cm}$ in height. The disc has two cylindrical slots, $40\ \mathrm{cm}$ apart ($r_\mathrm{1,2}^\mathrm{I}=20\ \mathrm{cm}$), which can hold different materials. Tungsten cylinders, $16.1\ \mathrm{cm}$ in diameter, can serve as rotating masses, with $m_\mathrm{1,2}^\mathrm{I}=(\rho_{W}-\rho_{Ti})V=30\ \mathrm{kg}$ effective mass each, where $\rho_{W}$ and $\rho_{Ti}$ are mass densities of tungsten and titanium respectively, and $V$ is the volume of a cylinder.

In the ideal case of symmetrical and matched DFGs, no effect on the TM is observed at $2 f_\mathrm{0}$ in the absence of $1/r^2$ violations. In the classical sense, this condition is satisfied if the DFG radii and placement follow the

\begin{eqnarray}
\label{etarelations}
r_\mathrm{1,2}^\mathrm{II}  &=& \sqrt{\eta} ~
r_\mathrm{1,2}^\mathrm{I} \\ \nonumber d^\mathrm{II}  &=&
\sqrt{\eta} ~ d^\mathrm{I}
\end{eqnarray}

scaling laws, where indices $\mathrm{I}$ and $\mathrm{II}$ are associated with the two DFGs, and $\eta = m_\mathrm{1,2}^\mathrm{II}/m_\mathrm{1,2}^\mathrm{I}$ is the mass ratio (see Fig.~\ref{tworotorD}.). Note, that if we take into account the cylindrical geometry of the TM, the scaling laws presented in Eq.~(\ref{etarelations}) will differ by O$(0.1\%)$, where the corresponding $r^\mathrm{II}_{1,2}$ and $d^\mathrm{II}$ will still be analytically calculable.

\begin{figure}
\begin{center}
  \includegraphics[width=3.4in]{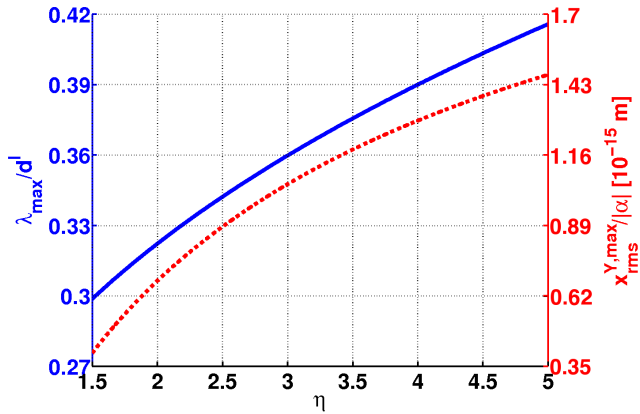}\\
  \caption{Simulation results for two DFGs positioned in a null experiment. The continuous line represents the plot
  of $\lambda_\mathrm{max}(\eta)$ expressed in $d^\mathrm{I}$ units, where $\lambda_\mathrm{max}$
  denotes the Yukawa interaction range, for which the $2 f_\mathrm{0}$ RMS displacement due to the Yukawa term is
  maximal. The corresponding RMS displacement of the TM (using $d^{\mathrm{I}}=2~\mathrm{m}$, $r^{\mathrm{I}}=20~\mathrm{cm}$, $m^{\mathrm{I}}=30~\mathrm{kg}$, and $f_0=5~\mathrm{Hz}$), expressed in
  $\left| \alpha  \right|$ units, is shown by the dashed curve.}
  \label{eta}
\end{center}
\end{figure}

In non-Newtonian dynamics, using the Yukawa formalism, limits on the parameter alpha are given as a function of the Yukawa range, lambda. First we address an ideal case, where the parameters of both DFGs are known exactly. As an example, let's assume the first DFG, with parameters described above, positioned at $d^\mathrm{I}=2\ \mathrm{m}$ from the center of mass of the IFO TM. Specifications and position for the second DFG are determined by Eq.~(\ref{etarelations}). On Fig.~\ref{eta}. we plot the interaction range for which the TM displacement at $2 f_\mathrm{0}$ is maximal, $\lambda_{\mathrm{max}}$, as a function of $\eta$ (continuous line). On the same plot, the dashed line shows the corresponding TM RMS displacement at $2 f_\mathrm{0}=10~\mathrm{Hz}$ due to the Yukawa term.

Using the above case and $\eta=2$, Fig.~\ref{eta}. gives an interaction range of $\lambda_\mathrm{max} = 0.64\ \mathrm{m}$, corresponding to a maximum displacement of the TM due to the Yukawa term with an RMS value of $x_\mathrm{rms} = ( 6.8 \times 10^{-16}\ \mathrm{m}) \times \left| \alpha \right|\ $. For the case of the AEI 10m detector with a noise floor of $\tilde{n} = 1.5 \times 10^{-17}\ \mathrm{m}/\sqrt{\mathrm{Hz}}$ at $10\ \mathrm{Hz}$, and an integration time of $T = 1$ day, a limit of $\left| \alpha \right| \simeq 2.3 \times 10^{-4}$ can be provided with a signal to noise ratio (defined as the ratio of the RMS signal to the displacement noise spectrum density integrated for a time $T$) of $\mathrm{SNR} = 3$, given the basic physical limitations described later in this section. In general terms, for an arbitrary noise floor $\tilde{n}$, and integration time $T$, the signal to noise ratio scales as

\begin{eqnarray}
\mathrm{SNR} = 3 \times \Bigg{(} \frac{1.5 \times 10^{-17}\
\mathrm{m}/\sqrt{\mathrm{Hz}}}{\tilde{n}}
\Bigg{)} \Bigg{(} \frac{T}{1\ \mathrm{day}} \Bigg{)}^{1/2} \hskip -0.1in &\times& \\
\nonumber && \hskip -1.8in \times \Bigg{(} \frac{x_\mathrm{rms}}{6.8
\times 10^{-16}\ \mathrm{m}} \Bigg{)} \Bigg{(} \frac{\left| \alpha
\right|}{2.3 \times 10^{-4}} \Bigg{)}
\end{eqnarray}

\subsection{Measurement Uncertainties}
\label{yukawaunc}
There are two fundamental classes of effects that limit our capability of measuring an inverse-square-law-violating gravitational force within the experimental concept. The first one is due to the noise level of the IFO, which effect can be suppressed by using sufficiently long integration times, as discussed in Sec.~\ref{yuk}. In this section we discuss some of the effects from the second class, which are due to the finite measuring and manufacturing precision when setting up the null-measurement.

Errors due to conceivable machining precision and procedure, uncertainties related to measurement of parameters of the final DFG products as well as their final placement relative to the TM and to each other will all infer an uncertainty in the induced displacement measurement, and propagate an error into the measurement of the non-Newtonian parameter $\alpha$. In case of purely symmetric DFGs positioned in a null experiment configuration, and taking into account only the Newtonian field component, the induced displacement of the TM will be dominated by a $4 f_\mathrm{0}$ term. In the presence of asymmetries, the displacement will also have components at the odd harmonics and the classical terms at $2 f_\mathrm{0}$ will not cancel each other completely. This means that the Newtonian peak due to this imperfect cancellation will appear at $2f_\mathrm{0}$, and will contaminate a potential Yukawa-signal from the very beginning of the measurement.

\begin{figure}[h]
\begin{center}
  \includegraphics[width=3.4in]{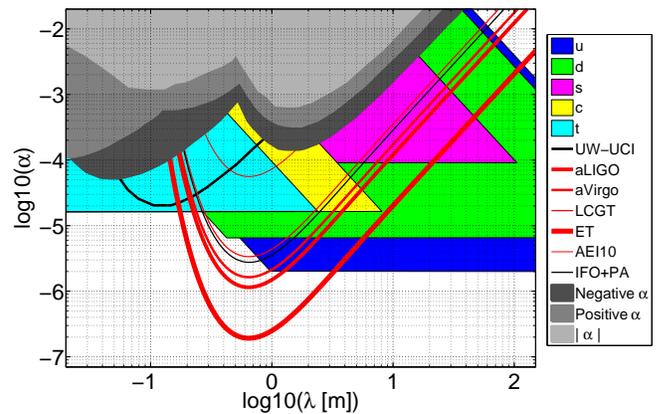}\\
  \caption{Predictions for the allowed lower limits on Yukawa parameter $|\alpha|$ via laser interferometric method proposed here. The grey shadings show the parameter regions excluded by previous measurements~\cite{AdelbergerReview2003,Irvine2006} that were sensitive to negative, positive, and absolute alpha values only. The thick black line corresponds to the limit achievable with an ongoing experiment, denoted by UW-UCI~\cite{Irvine2006}. The red curves show the results for various IFOs in case of infinite DFG machining precision, using an integration time of $T=10^7$ sec. The IFO+PA curve gives the 95\% confidence limits for all IFOs but LCGT and AEI 10m, when allowing DFG parameter uncertainties given in Sec.~\ref{feasibility}. The limiting factor in the measurement is the finite integration time when using LCGT and AEI 10m, and the finite machining precision in case of all the other IFOs. The coloured background areas correspond to different theoretical models presented in~\cite{NewmanBergBoynton2009}, that predict Yukawa parameters within the areas. Our tolerable parameter uncertainties were chosen such as to give the best coverage of the parallelogram corresponding to model "t", while taking into account the feasible parameter adjustment precision.}
  \label{perf}
\end{center}
\end{figure}

First, we calculate the achievable lower limits on the Yukawa strength parameter, $|\alpha|$, in case of ideal DFGs with no parameter errors. We chose the parameters of the first DFG to be $m^\mathrm{I} = 30\ \mathrm{kg}$, $d^\mathrm{I} = 2\ \mathrm{m}$, and $r^\mathrm{I} = 0.2\ \mathrm{m}$, and maximized the integration time of the measurement at $T = 10^{7}$ s ($\simeq 4$ months). Because of the slight dependence of the TM response on $\eta$ (see Fig.~\ref{eta}.), and taking into account manufactural limitations, we set $\eta$ to 2. We chose a conservative limit of $\mathrm{SNR}=8$ as the condition for detection. The optimal DFG frequencies that maximize the SNR in terms of $|\alpha|$ for the different detectors (see Fig.~\ref{SensCurves}), along with the lowest achievable limits on the strength parameter, $|\alpha^{*}|$, are given in Table~\ref{classicerrors}. Results on $|\alpha|$ as a function of the Yukawa scale parameter, $\lambda$, for five different IFOs are provided in Fig.~\ref{perf}. Note, that these results are also valid for the case, when the measurement setup errors are kept low enough to allow the detector noise levels to be the main limiting factors.

\begin{table}[h]
\centering
\caption{Results of the numerical calculations, excluding DFG-related experimental uncertainties, for achieving absolute limits on $|\alpha|$ in case of maximal detector response due to the Yukawa term at $2 f_\mathrm{0}$ ($|\alpha^*|$), for an integration time of $T = 10^{7}\ \mathrm{s} \simeq 4$ months. The DFG parameters used for generating the results are $m^{\mathrm{I}}=30$~kg, $d^{\mathrm{I}}=2$~m and $r^{\mathrm{I}}~=~0.2$~m. $\lambda_\mathrm{max}$ corresponding to $|\alpha^*|$ is $\lambda_\mathrm{max} = 0.64\ \mathrm{m}$ in all five cases. The results on optimal DFG frequencies for ET and AEI 10m are limited by the uncertainties of the IFO sensitivity models at low ($\lesssim10~\mathrm{Hz}$) frequencies.}
\resizebox{3.4in}{!}{
\begin{tabular}{c c c c c c}
	\hline \hline
	$T=10^{7}$ s & aLIGO & aVIRGO & ET & LCGT & AEI 10m \\
	\hline
	$f_0$ & $11$ Hz & $20$ Hz & $\lesssim5$ Hz & $26$ Hz & $\lesssim5$ Hz\\
	$|\alpha^*|$ & $1.2 \times 10^{-6}$ & $1.6 \times 10^{-6}$ & $1.9 \times 10^{-7}$ & $3.4 \times 10^{-6}$ & $5.6 \times 10^{-5}$ \\
	\hline \hline
\end{tabular}
}
\label{classicerrors}
\end{table}

In order to study the feasibility and real-life limitations of null-experiment concept, we numerically studied the effect of uncertainties associated with the system parameters included in the simulations (rotation frequency ($f_\mathrm{0}$), mass ($m$) and radius ($r$) of each DFG, as well as the DFG positions relative to the TM and to each other ($d$ and $\beta$)), thereby mimicking a realistic construction and measurement procedure. To do this, a large number ($N$) of hypothetical setups were Monte Carlo simulated with the DFG parameters normally distributed around a mean. The mean of the parameter distributions (e.g. $\left\langle m^{\mathrm{I}}\right\rangle = 30$~kg, $\left\langle d^{\mathrm{I}}\right\rangle =2\ \mathrm{m}$ and $\left\langle r^{\mathrm{I}}\right\rangle$~=~0.2~m) are chosen such as to maximize the response of the IFO in terms of $|\alpha|$, while taking into account the technically achievable range of values for each parameter, as well as safety issues.

We again set $\eta$ to $2$ for all setups, similarly to the ideal case. Following Eq.~(\ref{etarelations}), $\eta$ in this case determines the means of the distributions corresponding to parameters of the second DFG. The means of DFG operational frequencies were chosen to be equal to the values given in Table~\ref{classicerrors}. The low optimal DFG frequencies allow the effective masses of DFG fillings for the first DFG (rotating closer to the IFOs' vacuum chamber) to $m^{\mathrm{I}}=30\ \mathrm{kg}$. The highest corresponding mechanical energy of a DFG ($m^{\mathrm{II}}=60\ \mathrm{kg}$; $r^{\mathrm{II}}=\sqrt{2}\times0.2\ \mathrm{m}$; $f_0=26\ \mathrm{Hz}$) would still be less than $0.02\%$ of the energy proposed in a similar experiment described in~\cite{ShapiroPaper}. Note, that as an alternative to the DFG geometry we proposed in Sec.~\ref{yuk}, a rotating mass design similar to the one suggested by~\cite{ShapiroPaper} can also be used in the measurement.

Naturally, we are seeking to achieve the highest precision measurement currently possible or feasible in the near future, and we let the parameter uncertainties vary to values corresponding to state-of-the-art limits with the investment of plausible amount of work and finances. Our goal is to find, using Monte Carlo simulations, the limits on $|\alpha|$, allowed by using a two-DFG configuration with parameter uncertainties.

Each parameter associated to hypothetical setups were randomized independently, from a Gaussian distribution, using the same variances for corresponding parameters of different setups. We used a chosen set of variances for the parameter distributions (discussed in Sec.~\ref{feasibility}) that are within the limits of current state-of-the-art machining and measurement technologies, and at the same time, minimize the Newtonian peak we get at $2f_{0}$ due to imperfect cancellation. We generated $N=1000$ two-DFG setups, and determined the 95\% quantile of the induced Newtonian displacement values at $2f_{0}$ (from now on, denoted by $x^{N}_{95}$). Note that this quantile $x^{N}_{95}$ is completely independent from the detector noise and thus from the measurement integration time. We then, for each $\lambda$, calculated the $|\alpha|$ value for which the resulting Yukawa peak at $2f_{0}$ equals $x^{N}_{95}$. The limits on $|\alpha|$ defined this way, versus $\lambda$, are shown in Fig.~\ref{perf}.

Using a two-DFG setup with allowed parameter uncertainties given in Sec.~\ref{feasibility}, the limiting factor in case of the LCGT and AEI 10m IFO is its relatively high noise level at $2f_{0}$. However this means, that we can tolerate higher parameter uncertainties compared to the ones given in Sec.~\ref{feasibility}, when using LCGT and AEI 10m and an integration time of 4 months. In case of the other three IFOs (aLIGO, aVIRGO, and ET) the limiting factor is the imperfect cancellation of the Newtonian peaks at $2f_{0}$ due to parameter errors, when considering a 4 months long integration time. Therefore in order to reach the best possible $|\alpha|$ limits for the chosen set of parameter uncertainties, we can use a reduced integration time of O($10^{4} - 10^{6}~\mathrm{s}$), depending on which IFO we choose. The reach of the allowed exclusion region suggests that the displacement sensitivity of any dedicated IFO for the null-measurement should have comparable displacement sensitivity to advanced interferometric gravitational wave detectors. Custom configured interferometry could possibly have several advantages such as double DFG compatible suspensions, being able to reduce the distance between DFGs and the IFO test mass, and thus to obtain better measurement results.

\subsection{Measurement feasibility}
\label{feasibility}

Monte Carlo test results presented in Sec.~\ref{yukawaunc} were provided by using a chosen set of DFG parameter uncertainties, that gives the best coverage of Yukawa parameter predictions given by the "t" model (see~\cite{NewmanBergBoynton2009} for more details), and that are still within the feasible measuring and machining precision. An optimization of tolerable DFG parameter uncertainties in terms of measuring an ISL-violating gravitational force can be given with the Fisher matrix method \cite{fisher}, however this analysis is beyond the scope of this proof-of-concept paper. In this section we discuss the DFG parameter uncertainties that correspond to the IFO+PA curve given in Fig.~\ref{perf}. Note, that we assumed that the parameter uncertainties are kept at the same levels throughout the whole integration time of the Yukawa measurement. We also discuss the possible methods in measuring and machining that allow us to reach the chosen DFG parameter precision.

In the conceptual experimental setup considered the two DFGs are rotating at $f_{\mathrm{0}}=$ 11 Hz, 20 Hz, 5 Hz, 26 Hz, and 5 Hz using aLIGO, aVIRGO, ET, LCGT, and AEI 10m, respectively. In all cases the two-DFGs in a setup have to be in a relative phase of $\beta = \pi / 2$. In our simulations we chose the following measurement uncertainties:

\begin{itemize}

	\item Uncertainty in the frequency of the first DFG (absolute measurement): $\delta f_{1}~=~10^{-6}\ \mathrm{Hz}$

	\item Uncertainty in the frequency of the second DFG, relative to the first DFG (relative measurement): $\delta f_{12}~=~10^{-9}\ \mathrm{Hz}$

	\item Uncertainty in the initial phase difference between the two DFGs (absolute measurement): $\delta \beta~=~10^{-8}\ \mathrm{rad}$
	
\end{itemize}

Frequencies and the relative phase between the different DFGs can be finetuned to their desired value by calibrating and locking the DFGs' rotation signal with an accurate and low-phase-noise clock (e.g., good Cesium or GPS), also used for time-stamping the IFO data. A sophisticated DFG frequency (and phase) control system can be constructed allowing nanosecond timing precision.

In our simulations we used tungsten filling masses of 30 kg for the first, and $\eta \times 30$ kg for the second DFG. The masses had the following permissible uncertainties:

\begin{itemize}

	\item Uncertainty in the exact mass of one of the tungsten cylinders in the first DFG (absolute measurement): $\delta m_{1}=5 \times 10^{-5}\ \mathrm{kg}$

	\item Uncertainty in the mass of any other DFG cylinders of the two-DFG setup (relative measurement). The mass of one of the tungsten cylinders built into the second DFG has to be $\eta$ times the mass of the one in the first DFG, where we proposed to use $\eta=2$. This way we can use a relative measurement, and thus, we have to adjust the mass of one of the cylinders in the second DFG to the sum of the masses in the first DFG. We adjust the mass of the second tungsten cylinder in the second DFG to the mass of the first cylinder in the same DFG. The mass of the second cylinder in the second DFG can be then finetuned to the mass of the first cylinder within the same DFG. The tolerable uncertainty of these relative measurements and mass adjustments: $\delta m_{12}=10^{-6}\ \mathrm{kg}$

\end{itemize}

For the DFG mass measurements, one can use scales such as the Sartorius ME415S scale~\cite{sartoriusme415s}, that can determine the absolute value of the masses with an uncertainty of $\sim10^{-7}$ kg with 410 g weighing capacity. Using state-of-the-art mass comparators allows us to reduce the relative measurement (mass-to-mass comparison) error to $\sim2\times 10^{-9}$ kilograms (e.g~\cite{balance3}). A complete tungsten cylinder can be built from an assembly of several smaller disks, allowing us to reduce uncertainties with precision scales having lower weighing capacity. These smaller and lighter components of the tungsten fillings are also easier to be manufactured and handled. A combination of precision absolute scales, mass comparators, and fine mass adjustment through abrasion leads to precise mass standards and also allows manufacturing of matched mass pairs beyond the load limit of the absolute scale.

Arm lengths and DFG positions have to be set to definite values (in our example case $r_\mathrm{1,2}^\mathrm{I} = 0.2\ \mathrm{m}$, $r_\mathrm{1,2}^\mathrm{II} = \sqrt{2}r_\mathrm{1,2}^\mathrm{I}$, $d^\mathrm{I} = 2\ \mathrm{m}$, and $d^\mathrm{II} = \sqrt{2}d^\mathrm{I}$) to achieve precise cancellation of the Newtonian component. The corresponding parameters have the following uncertainties:

\begin{itemize}

	\item Uncertainty in the distance between the first DFG's rotation center and the TM's center of gravity ($d^\mathrm{I}$ absolute measurement): $\delta d_{1}~=~10^{-2}\ \mathrm{m}$

	\item Uncertainty in the distance between the two DFGs relative to $d^\mathrm{I}$ (relative measurement): $\delta d_{12}~=~10^{-7}\ \mathrm{m}$
	
	\item Uncertainty in the length of one arm in the first DFG ($r^\mathrm{I}$ absolute measurement): $\delta r_{1}~=~10^{-3}\ \mathrm{m}$	

	\item Uncertainty of in the equality of arms within one DFG ($\delta r_{11,22}$), and the uncertainty in the length of one arm of the second DFG relative to $r^{I}$ ($\delta r_{12}$) (relative measurements): $\delta r_{11,22}~=~\delta r_{12}~=~3\times 10^{-8}\ \mathrm{m}$
					
\end{itemize}

Distance-like quantities ($r$-s and $d$-s) can be measured using interferometric methods with accuracy significantly surpassing the wavelength of the laser light. The exact technique of the measurement greatly depends on the DFG placement, accelerating and controlling configuration, and is beyond the scope of this article.

Setting the arm lengths within one DFG to be equal with a precision of $3\times10^{-8}$ m is challenging, but possibly feasible. The uncertainty of equivalence of arms within one DFG will be in the order of the equivalence of masses within the DFG, and the uncertainty of $\mathcal{M}_{1}$ being zero. The feasible precision of mass-to-mass comparison can be as low as $\sim2\times10^{-9}$ kg \cite{balance3}. In case of a non-zero $\mathcal{M}_1$, the axis of rotation of a spinning DFG will subject a periodic driving force (that is proportional to $\sim f_{0}^{2}$), causing the support of the axis to vibrate. The amplitude of the vibration can be measured with high-precision interferometry, and the placement of the center of rotation can be adjusted until the vibration amplitude is zero within the measurement precision. The adjustment can be carried out using much higher DFG frequencies than the values given in Table~\ref{classicerrors}, allowing an amplification of DFG vibrations in case of a non-zero $\mathcal{M}_1$.

The non-Newtonian and the classical gravitational potentials scale differently with $d$, the distance between the TM and the center of rotation of a DFG. Using the two DFGs positioned in a null experiment configuration, and measuring the TM displacement spectrum in case of different $d$-s, can also help us to determine measurement and setup inaccuracies other than errors of $d$.

It is possible that periodically varying one or more of the tunable parameters (e.g., distance or phase) will allow precise, $in-situ$ tuning of the cancellation necessary for the null-measurement. Also, an additional effect can be exploited to estimate the dominating error in the two-DFG setup parameters. By applying the parameter uncertainty tolerances given in this section, the TM's induced displacement at $f_{0}$ due to imperfect cancellation of the Newtonian term will be orders of magnitudes above the detector noise level and the Yukawa term at $f_{0}$, and thus will be detectable. By monitoring this signal at $f_{0}$ we can give a rough estimation on the dominating error in the parameters of the two-DFG setup.

We investigated the projected contributions of the different parameter uncertainties, relative to each other, to the uncertainty of an $|\alpha|$ measurement. To do this, we carried out a series of Monte Carlo simulations, where in each simulation, we set the relative uncertainty of a chosen parameter to $10^{-6}$, and kept the uncertainties of all the other parameters at zero level. We simulated $N=1000$ two-DFG setups, and calculated the relative error of $|\alpha|$ obtained from the $N=1000$ hypothetical measurements of $|\alpha|$. In order to characterize the relative dominance of the different parameters compared to each other, we normalized the measured relative errors of $|\alpha|$ with the highest such relative error value we obtained. This way, our final results were basically independent from the $10^{-6}$ relative uncertainty chosen for each individual setup parameters. The resulting relative dominances of the different parameter uncertainties in the $|\alpha|$ measurement are given in Table~\ref{alpharelativeerrors}. Our results show that the dominant parameter - for which the allowed uncertainty, $\delta d_{12}$, should be kept as low as possible -  is the distance between the two DFGs, relative to $d^{\mathrm{I}}$.

\begin{table}[h]
\centering
\caption{The relative dominance (RD) of parameter uncertainties in terms of their contribution to the uncertainty of $|\alpha|$ in a hypothetical measurement. We carried out $N=1000$ Monte Carlo trials for each uncertainties, setting the chosen uncertainty to a $10^{-6}$ relative value, and while setting all the other uncertainties to zero. The relative dominance for each parameter uncertainty was obtained by calculating the relative uncertainty of $|\alpha|$ measured from the $N$ trials, and normalizing it by the largest such relative uncertainty value (thus we get a relative dominance of 1 for the most dominant parameter uncertainty).}

\begin{tabular}{c c c c c c}
	\hline \hline
	
	Uncertainty & $\delta d_{12}$ & $\delta d_{1}$ & $\delta f_{12}$ & $\delta r_{12}$ & $\delta r_{11,22}$ \\
	\hline
	RD & $ 1.0 $ & $ 0.52 $ & $ 0.37 $ & $ 0.32 $ & $ 0.28 $ \\

	\hline \hline

	Uncertainty & $\delta f_{1}$ & $\delta r_{1}$ & $\delta m_{12}$ & $\delta m_{1}$ & $\delta \beta$ \\
	\hline
	RD & $ 0.26 $ & $ 0.26 $ & $ 0.217 $ & $ 0.127 $ & $ 2.3\times10^{-6} $\\

	\hline \hline
\end{tabular}

\label{alpharelativeerrors}
\end{table}

In this subsection we addressed the fundamental sources of measurement errors and the uncertainty limits for each parameter of interest. We also pointed out, that uncertainty values for DFG parameters used in our proposed experimental setup are within the limits of current state-of-the-art machining and measurement technologies. For a more realistic application a full dynamic finite element simulation of the mass distributions must be performed. The expansion and stress factors of the DFGs under prolonged operating conditions must be modeled and simulated, then subsequently measured and taken into account. Practical limitations due to the complex mechanical and detector dynamics/geometry around the test mass and other second order error sources need to be investigated, via experiments and more sophisticated simulations as they might prevent us from reaching the full potential of DFG device based measurements. For example, in practice material inhomogeneities can give rise to a substantial uncertainty that could be minimized by using NDT-rated materials~\cite{NDT} and also reduced by different geometries. Alternative, more simplified DFG geometries different from the one proposed here (e.g. rotating rods) could also possibly lead to higher measuring and machining precision, and therefore should be studied carefully. All these issues will be the scopes of future studies.

\section{Safety}
\label{safe}
Significant kinetic energy (i.e. $\sim$~100 kJ) is stored in the DFG once it rotates, therefore crucial safety considerations must be addressed. There are two major points of failure management to be concerned with.

(a.) The vacuum chamber of the DFGs must be made strong enough to withstand the damage of an accidentally disintegrating disk. This is the standard solution for high speed gyroscopes.

(b.) For added security, the gap between the inner wall of the vacuum chamber and the outer edge of the rotating disk could be kept relatively {\it small}. In the event of an incident where the DFG's material starts to yield or its angular acceleration is uncontrolled the disk will expand radially touching the sidewall and may slowly stop, potentially preempting some of the catastrophic failure modes. Adding a heavy ring in contact with the inner lateral wall of the vacuum chamber (similarly as described in~\cite{Astone1998}) would absorb a huge angular momentum of the DFG and dissipate its rotation energy via the friction against the lateral wall.

These conditions can be met using Finite Element Analysis (FEA) aided design, in-house destructive testing of sacrificial parts and relying only on the best base materials.

\section{Conclusion}
Two DFGs in a null experiment setup in conjunction with an interferometric sensor allow for studies of composition independent non-Newtonian contribution to the classical gravitational field in the meter scale. Simulation results presented in this paper indicate that by taking advantage of two matched DFGs and the relatively large bandwidth and high sensitivity of state-of-the-art interferometric sensors there is an exciting opportunity to explore $\alpha$ below the current limit in the $\lambda \simeq 0.1-10$ meter Yukawa range addressing standing theoretical predictions.

For the proof of concept study, we chose a conservative 2 meter distance for our device from the test mass of the interferometric detector and used the sensitivity of advanced and third-generation gravitational wave detectors as the realistic baseline. We intend to note here, that putting the devices considerably closer to the test mass can yield orders of magnitude better limits in ISL violation parameters through the characteristic length scale changes for the Yukawa case.

There are many practical details that still need close attention when designing and manufacturing a practical device. Finite element analysis of the DFGs and subsequent experimental studies are necessary to completely understand the stresses the DFG is subjected to. The DFGs need to be enclosed in separate vacuum chambers seated on seismic attenuation stage(s). Prototype design and test will be necessary to balance the disk and test vibration control. Direct gravitational coupling into the test mass suspension and seismic isolation system must be carefully designed, simulated and mitigated. Any structure supporting the test mass must be placed far away (i.e., tens of meters) from the double DFG assembly. Other mostly practical problems, such as safety, can also be solved as was shown in past applications/experiments that have used rapidly rotating instruments (for examples see references in Sec.~\ref{introduction}). It is also likely that dedicated and custom designed interferometric sensors will be necessary and the direct use of advanced interferometric gravitational wave detectors will not be ideal for technical reasons.

In spite of the promising estimates one should not underestimate the technical difficulties imposed by (1) the required design precision, (2) the manufacturing of the DFGs and (3) the unknown, but experimentally possible to study, complex nature of the coupling routes of the dynamic gravity fields of the DFGs into the complicated mechanical structure of the IFO test masses.

\section{Acknowledgements}

The authors are grateful for the support of the United States National Science Foundation under cooperative agreement PHY-04-57528 and Columbia University in the City of New York. We acknowledge the support of the Hungarian National Office for Research and Technology (NKTH) through the Polanyi program (grant no. KFKT-2006-01-0012). We greatly appreciate the support of LIGO collaboration. We are indebted to many of our colleagues for frequent and fruitful discussion. In particular we'd like to thank M.~Landry, P.~Sutton, D.~Sigg, G.~Giordano, R.~Adhikari, V.~Sandberg, P.~Shawhan, R.~DeSalvo, H.~Yamamoto, J. Harms, C.~Matone and Z.~Frei for their support and valuable comments on the manuscript.


\begin{thebibliography}{28}
\expandafter\ifx\csname natexlab\endcsname\relax\def\natexlab#1{#1}\fi
\expandafter\ifx\csname bibnamefont\endcsname\relax
  \def\bibnamefont#1{#1}\fi
\expandafter\ifx\csname bibfnamefont\endcsname\relax
  \def\bibfnamefont#1{#1}\fi
\expandafter\ifx\csname citenamefont\endcsname\relax
  \def\citenamefont#1{#1}\fi
\expandafter\ifx\csname url\endcsname\relax
  \def\url#1{\texttt{#1}}\fi
\expandafter\ifx\csname urlprefix\endcsname\relax\def\urlprefix{URL }\fi
\providecommand{\bibinfo}[2]{#2}
\providecommand{\eprint}[2][]{\url{#2}}

\bibitem[{adl()}]{adligohomepage}
\bibinfo{note}{{\url{http://www.ligo.caltech.edu/advLIGO}}}.

\bibitem[{adv()}]{advirgohomepage}
\bibinfo{note}{{\url{http://wwwcascina.virgo.infn.it/advirgo/}}}.

\bibitem[{\citenamefont{{{Kuroda}, K. and the LCGT
  Collaboration}}(2010)}]{LCGTOverview}
\bibinfo{author}{\bibnamefont{{{Kuroda}, K. and the LCGT Collaboration}}},
  \bibinfo{journal}{Classical and Quantum Gravity}
  \textbf{\bibinfo{volume}{27}}, \bibinfo{pages}{084004}
  (\bibinfo{year}{2010}).

\bibitem[{\citenamefont{{{Gossler}, S., {et al}}}(2010)}]{AEI10m}
\bibinfo{author}{\bibnamefont{{{Gossler}, S., {et al}}}},
  \bibinfo{journal}{Class. Quantum Grav.} \textbf{\bibinfo{volume}{27}},
  \bibinfo{pages}{084023} (\bibinfo{year}{2010}).

\bibitem[{\citenamefont{{Chassande-Mottin, E., Hendry, M., Sutton, P. J.,
  M\'arka, S.}}(2010)}]{ChassandeMottinetal}
\bibinfo{author}{\bibnamefont{{Chassande-Mottin, E., Hendry, M., Sutton, P. J.,
  M\'arka, S.}}}, \bibinfo{journal}{to appear in Special issue of GRG on the
  Einstein Telescope}  (\bibinfo{year}{2010}), \eprint{arXiv:1004.1964}.

\bibitem[{\citenamefont{{Hughes, S. A., M\'arka, S., Bender, P. L. and Hogan,
  C. J.}}(2001)}]{snomass2001}
\bibinfo{author}{\bibnamefont{{Hughes, S. A., M\'arka, S., Bender, P. L. and
  Hogan, C. J.}}}, \bibinfo{journal}{{Proc. of the APS/DPF/DPB Summer Study on
  the Future of Particle Physics (Snowmass 2001) ed. Graf, N. eConf}, {
  {C010630}}, P402., { arXiv:astro-ph/0110349}}  (\bibinfo{year}{2001}).

\bibitem[{\citenamefont{{{Matone}, L., {Raffai}, P., {M\'arka}, S., {Grossman},
  R., {Kalmus}, P., {M\'arka}, Z., {Rollins}, J. and {Sannibale},
  V.}}(2007)}]{rotorCQG}
\bibinfo{author}{\bibnamefont{{{Matone}, L., {Raffai}, P., {M\'arka}, S.,
  {Grossman}, R., {Kalmus}, P., {M\'arka}, Z., {Rollins}, J. and {Sannibale},
  V.}}}, \bibinfo{journal}{Class. Quantum Grav.} \textbf{\bibinfo{volume}{24}},
  \bibinfo{pages}{2217} (\bibinfo{year}{2007}), \eprint{gr-qc/0701134}.

\bibitem[{\citenamefont{{{Forward}, R. L. and {Miller}, L.
  R.}}(1967)}]{ForwardMiller1967}
\bibinfo{author}{\bibnamefont{{{Forward}, R. L. and {Miller}, L. R.}}},
  \bibinfo{journal}{J. Appl. Phys.} \textbf{\bibinfo{volume}{38}},
  \bibinfo{pages}{512} (\bibinfo{year}{1967}).

\bibitem[{\citenamefont{{{Sinsky}, J. and {Weber},
  J.}}(1967)}]{SinskyWeber1967}
\bibinfo{author}{\bibnamefont{{{Sinsky}, J. and {Weber}, J.}}},
  \bibinfo{journal}{Physical Review Letters} \textbf{\bibinfo{volume}{18}},
  \bibinfo{pages}{795} (\bibinfo{year}{1967}).

\bibitem[{\citenamefont{{{Sinsky}, J.~A.}}(1968)}]{Sinsky1968}
\bibinfo{author}{\bibnamefont{{{Sinsky}, J.~A.}}}, \bibinfo{journal}{Physical
  Review} \textbf{\bibinfo{volume}{167}}, \bibinfo{pages}{1145}
  (\bibinfo{year}{1968}).

\bibitem[{\citenamefont{{{Hirakawa}, H., {Tsubono}, K. and {Oide},
  K.}}(1980)}]{Hirakawa1980}
\bibinfo{author}{\bibnamefont{{{Hirakawa}, H., {Tsubono}, K. and {Oide}, K.}}},
  \bibinfo{journal}{Nature} \textbf{\bibinfo{volume}{283}},
  \bibinfo{pages}{184} (\bibinfo{year}{1980}).

\bibitem[{\citenamefont{{{Oide}, K., {Tsubono}, K. and {Hirakawa},
  H.}}(1980)}]{oide}
\bibinfo{author}{\bibnamefont{{{Oide}, K., {Tsubono}, K. and {Hirakawa}, H.}}},
  \bibinfo{journal}{Jap. J. Appl. Phys.} \textbf{\bibinfo{volume}{19}},
  \bibinfo{pages}{L123} (\bibinfo{year}{1980}).

\bibitem[{\citenamefont{{{Suzuki}, T., {Tsubono}, K. and {Kuroda},
  K.}}(1981)}]{suzuki}
\bibinfo{author}{\bibnamefont{{{Suzuki}, T., {Tsubono}, K. and {Kuroda}, K.}}},
  \bibinfo{journal}{Jap. J. Appl. Phys.} \textbf{\bibinfo{volume}{20}},
  \bibinfo{pages}{L498} (\bibinfo{year}{1981}).

\bibitem[{\citenamefont{{{Ogawa}, Y., {Tsubono}, K. and {Hirakawa},
  H.}}(1982)}]{Ogawa1982}
\bibinfo{author}{\bibnamefont{{{Ogawa}, Y., {Tsubono}, K. and {Hirakawa},
  H.}}}, \bibinfo{journal}{Phys. Rev. D.} \textbf{\bibinfo{volume}{26}},
  \bibinfo{pages}{729} (\bibinfo{year}{1982}).

\bibitem[{\citenamefont{{{Kuroda}, K. and {Hirakawa}, H.}}(1985)}]{Kuroda1985}
\bibinfo{author}{\bibnamefont{{{Kuroda}, K. and {Hirakawa}, H.}}},
  \bibinfo{journal}{Phys. Rev. D.} \textbf{\bibinfo{volume}{32}},
  \bibinfo{pages}{342} (\bibinfo{year}{1985}).

\bibitem[{\citenamefont{{{Astone}, P., {et al.}}}(1991)}]{Astone1991}
\bibinfo{author}{\bibnamefont{{{Astone}, P., {et al.}}}}, \bibinfo{journal}{Z.
  Phys. C} \textbf{\bibinfo{volume}{50}}, \bibinfo{pages}{21}
  (\bibinfo{year}{1991}).

\bibitem[{\citenamefont{{{Astone}, P., {et al.}}}(1998)}]{Astone1998}
\bibinfo{author}{\bibnamefont{{{Astone}, P., {et al.}}}},
  \bibinfo{journal}{European Physical Journal C} \textbf{\bibinfo{volume}{5}},
  \bibinfo{pages}{651} (\bibinfo{year}{1998}).

\bibitem[{\citenamefont{{{Adelberger}, E.~G., {Heckel}, B.~R. and {Nelson},
  A.~E.}}(2003)}]{AdelbergerReview2003}
\bibinfo{author}{\bibnamefont{{{Adelberger}, E.~G., {Heckel}, B.~R. and
  {Nelson}, A.~E.}}}, \bibinfo{journal}{Annual Review of Nuclear and Particle
  Science} \textbf{\bibinfo{volume}{53}}, \bibinfo{pages}{77}
  (\bibinfo{year}{2003}), \eprint{hep-ph/0307284}.

\bibitem[{\citenamefont{{{Hoskins}, J.~K., {Newman}, R.~D., {Spero}, R. and
  {Schultz}, J.}}(1985)}]{Hoskins1985}
\bibinfo{author}{\bibnamefont{{{Hoskins}, J.~K., {Newman}, R.~D., {Spero}, R.
  and {Schultz}, J.}}}, \bibinfo{journal}{Phys. Rev. D.}
  \textbf{\bibinfo{volume}{32}}, \bibinfo{pages}{3084} (\bibinfo{year}{1985}).

\bibitem[{\citenamefont{{{Moody}, M.~V. and {Paik}, H.~J.}}(1993)}]{MoodyPaik}
\bibinfo{author}{\bibnamefont{{{Moody}, M.~V. and {Paik}, H.~J.}}},
  \bibinfo{journal}{Physical Review Letters} \textbf{\bibinfo{volume}{70}},
  \bibinfo{pages}{1195} (\bibinfo{year}{1993}).

\bibitem[{\citenamefont{{{Newman}, R., {Berg}, E. and {Boynton},
  P.}}(2009)}]{NewmanBergBoynton2009}
\bibinfo{author}{\bibnamefont{{{Newman}, R., {Berg}, E. and {Boynton}, P.}}},
  \bibinfo{journal}{Space Science Reviews} \textbf{\bibinfo{volume}{148}},
  \bibinfo{pages}{175} (\bibinfo{year}{2009}).

\bibitem[{\citenamefont{{{Kawamura}, S., {Hazel}, J. and {Raab},
  F.}}(1996)}]{TMsuspension}
\bibinfo{author}{\bibnamefont{{{Kawamura}, S., {Hazel}, J. and {Raab}, F.}}}
  (\bibinfo{year}{1996}), \bibinfo{note}{{LIGO Internal Document T960074}}.

\bibitem[{\citenamefont{{{Ballmer}, S., {M{\'a}rka}, S. and {Shawhan},
  P.}}(2010)}]{ShapiroPaper}
\bibinfo{author}{\bibnamefont{{{Ballmer}, S., {M{\'a}rka}, S. and {Shawhan},
  P.}}}, \bibinfo{journal}{Class. Quantum Grav.} \textbf{\bibinfo{volume}{27}},
  \bibinfo{pages}{185018} (\bibinfo{year}{2010}).

\bibitem[{\citenamefont{{{Boynton}, P.~E., {Bonicalzi}, R.~M., {Kalet}, A.~M.,
  {Kleczewski}, A.~M., {Lingwood}, J.~K., {McKenney}, K.~J., {Moore}, M.~W.,
  {Steffen}, J.~H., {Berg}, E.~C., {Cross}, W.~D., {Newman}, R.~D. and
  {Gephart}, R.~E.}}(2007)}]{Irvine2006}
\bibinfo{author}{\bibnamefont{{{Boynton}, P.~E., {Bonicalzi}, R.~M., {Kalet},
  A.~M., {Kleczewski}, A.~M., {Lingwood}, J.~K., {McKenney}, K.~J., {Moore},
  M.~W., {Steffen}, J.~H., {Berg}, E.~C., {Cross}, W.~D., {Newman}, R.~D. and
  {Gephart}, R.~E.}}}, \bibinfo{journal}{New Astronomy Reviews}
  \textbf{\bibinfo{volume}{51}}, \bibinfo{pages}{334} (\bibinfo{year}{2007}).

\bibitem[{\citenamefont{{{Spall}, J. C.}}(2005)}]{fisher}
\bibinfo{author}{\bibnamefont{{{Spall}, J. C.}}}, \bibinfo{journal}{Journal of
  Computational and Graphical Statistics} \textbf{\bibinfo{volume}{14}},
  \bibinfo{pages}{889} (\bibinfo{year}{2005}).

\bibitem[{sar()}]{sartoriusme415s}
\bibinfo{note}{\url{http://www.dataweigh.com/products/product_detail.asp?Produ%
ctID=597}}.

\bibitem[{bal()}]{balance3}
\bibinfo{note}{{\url{http://sartorius.balances.com/sartorius/mass-comparators.%
html}}}.

\bibitem[{\citenamefont{{{Boving}, K. G.}}(1989)}]{NDT}
\bibinfo{author}{\bibnamefont{{{Boving}, K. G.}}}, \emph{\bibinfo{title}{{NDE
  Handbook, Non-Destructive Examination Methods for Condition Monitoring}}}
  (\bibinfo{publisher}{Teknisk Forlag A/S}, \bibinfo{year}{1989}).

\end{thebibliography}
\end{document}